\documentclass[12pt,preprint]{aastex}

\newcommand{\be}{\begin{equation}}
\newcommand{\ee}{\end{equation}}

\begin{document}

\title{The Extinction Toward the Galactic Bulge from RR Lyrae Stars}

\author{Andrea Kunder} 
\affil{E-mail: andrea.m.kunder@dartmouth.edu}
\affil{Dartmouth College, 6127 Wilder Lab, Hanover, NH 03755}
\author{Piotr Popowski} 
\affil{E-mail: popowski@mpa-garching.mpg.de}
\affil{Max-Planck-Institut f\"{u}r Astrophysik,
Karl-Schwarzschild-Str. 1, Postfach 1317, 85741 Garching bei
M\"{u}nchen, Germany}
\affil{E-mail: Andrea.M.Kunder@Dartmouth.edu and popowski@mpa-garching.mpg.de}
\author{Kem H. Cook}
\affil{IGPP/LLNL, P.O.Box 808, Livermore, CA 94551, USA}
\affil{E-mail: kcook@igpp.ucllnl.org}
\author{Brian Chaboyer} 
\affil{Dartmouth College, 6127 Wilder Lab, Hanover, NH 03755}
\affil{E-mail: chaboyer@heather.dartmouth.edu}
\begin{abstract}
We present mean reddenings toward 3525 RR0 Lyrae stars from the Galactic 
bulge fields of the MACHO Survey. These
reddenings are determined using the color at minimum $V$-band light of the 
RR0 Lyrae stars themselves and are found to be in general agreement 
with extinction estimates at the same location obtained 
from other methods.  Using 3256 stars located in the Galactic Bulge, 
we derive the selective extinction coefficient 
$R_{V,VR}=A_V/E(V-R) = 4.3 \pm 0.2$.  This value is what is expected for a
standard extinction law with $R_{V,BV} = 3.1 \pm 0.3$.

\end{abstract}

\section{Introduction}

Studies of stars in the central Galactic region are essential for
an understanding of the Milky Way.  For example, the questions of the 
structure and extent of the Galactic bar, the star formation history, 
and the distance to the center of our Galaxy can all be answered by 
analyzing various stellar tracers.  
Knowledge of the Galactic extinction is mandatory to begin 
investigations of these stars in a quantitative fashion.  
A good understanding of extinction is particularly important for the
direction toward the Galactic Bulge and within the Galactic plane, 
where extinction is very significant.

Extinction $A_\lambda$ changes as a function of wavelength $\lambda$.  
The shape of the extinction curve is characterized by the
quantity $R_V$, the ratio of total to selective extinction;  
\citet{car89}
have shown that the 
family of curves characterized by this single parameter provides a good
fitting approximation to most extinction curves.
Thus, if the value of $R_V$ can be determined 
(e.g., from optical and IR photometry), then one can approximately find
the properties of the entire UV-IR extinction curve.

There are a few suggestions that the coefficient of selective extinction in 
the direction of the Galactic Bulge may be non-standard 
\citep[e.g.][]{pop00}.  
Anomalous extinction, i.e., extinction described by smaller 
$R_{V,BV}$ or 
$R_{V,VR}$ ratio than the standard value of $R_{V,BV}$ = 3.1,
can significantly affect analysis of the stellar
populations in the Galactic Bulge.  Here, $R_{V,BV}$ is the
coefficient of selective extinction $A_V/E(B-V)$, 
with reference to the $B$ and $V$ filters, and 
$R_{V,VR} =  A_V/E(V-R)$, alluding to the $V$ and $R$ filters.
Especially with
the onset of a new generation of large telescopes, this problem
becomes increasingly pressing and important, and one that can be
addressed with this study.  Here, optical $V$ and $R$-band data for 
3256 RR0 Lyrae stars\footnote{RR0 
stars have been traditionally called RRab stars and 
are simply fundamental pulsators.  For the new, intuitive nomenclature, see 
\citet{alc00}.} from the MACHO microlensing experiment are used to find both 
the coefficient of selective extinction and to 
compute the absolute reddening near the Galactic 
bulge/bar.  The value of $R_{V,BV}$ is known to vary
in certain directions, and reddening laws are normally derived from
differential measurements of reddened and unreddened luminous early-type
stars.  However, as such stars are rare in the Galactic Bulge, the normal
derivation of $R_{V,BV}$ can not be carried out.
The $R_{V,BV}$ values determined from the RR Lyrae
stars are compared with other reddening relations.

The data used in this analysis are based on the MACHO photometry of 3674 
RR0 Lyrae stars toward the Galactic Bulge. Cook, Kunder, \& Popowski (2007,
in preparation).
Cook et al.\ (2007, in preparation) divide the MACHO bulge RR0 Lyrae 
variables into 
those associated with the Galactic Bulge (3256 stars), and the ones 
associated with the neighboring Sagittarius dwarf galaxy (409 stars).  
Their classification is assumed in this work.

The structure of this paper is the following.  The method and 
resultant coefficient of selective 
extinction toward the Bulge is discussed and compared to other estimates  
in \S 2.  Reddenings of the Bulge RR0
Lyrae stars are determined in \S 3 using their minimum V-band light colors
These reddenings are then compared to \citet{pop03} reddening 
estimates and the outliers are discussed.

\section{Coefficient of Selective Extinction}
\subsection{General approach}
As the bulge RR Lyrae stars span a much larger range in color than Sgr
ones, a coefficient of selective extinction is determined using only this 
subsample.  Assuming that the extinction affects the $V$ magnitudes 
much more 
significantly than the distance spread, the intrinsic scatter in 
absolute magnitudes, and the scatter due to various observational errors,
the slope of the stellar locus on a $V_{RR}$ (mean V magnitudes of the 
RR Lyrae stars) versus $(V-R)_{RR}$ (mean color of the 
RR Lyrae stars)
color-magnitude diagram (CMD) yields $R_{V,VR} \equiv A_V/E(V-R)$
\citep{kunderConf}.  
In other words, if in the distance modulus, 
\begin{equation}
V_{RR}-M_V-A_V = 5 \, \, \log(d)
\label{distance}
\end{equation}
we assume $M_V$ and $d$ is approximately constant, then the apparent
magnitude, $V_{RR}$, is proportional to $A_V$:
\begin{equation}
V_{RR} \sim A_V \sim \frac{E(V-R)}{R_{V,VR}}.
\end{equation}
Taking the $(V-R)_0$ of each RR0 Lyrae star to be approximately constant,
\begin{equation}
V_{RR} \sim \frac{(V-R)_{RR}}{R_{V,VR}}
\end{equation}
which is in the form of a general linear equation with $V_{RR}$ as the 
independent variable,
$(V-R)_{RR}$ as the dependent variable, and $R_{V,VR}$ as the slope.
It is shown in the following section that the above conditions are 
approximately met. 

\begin{figure}
\epsscale{1.0}
\includegraphics[width=16cm]{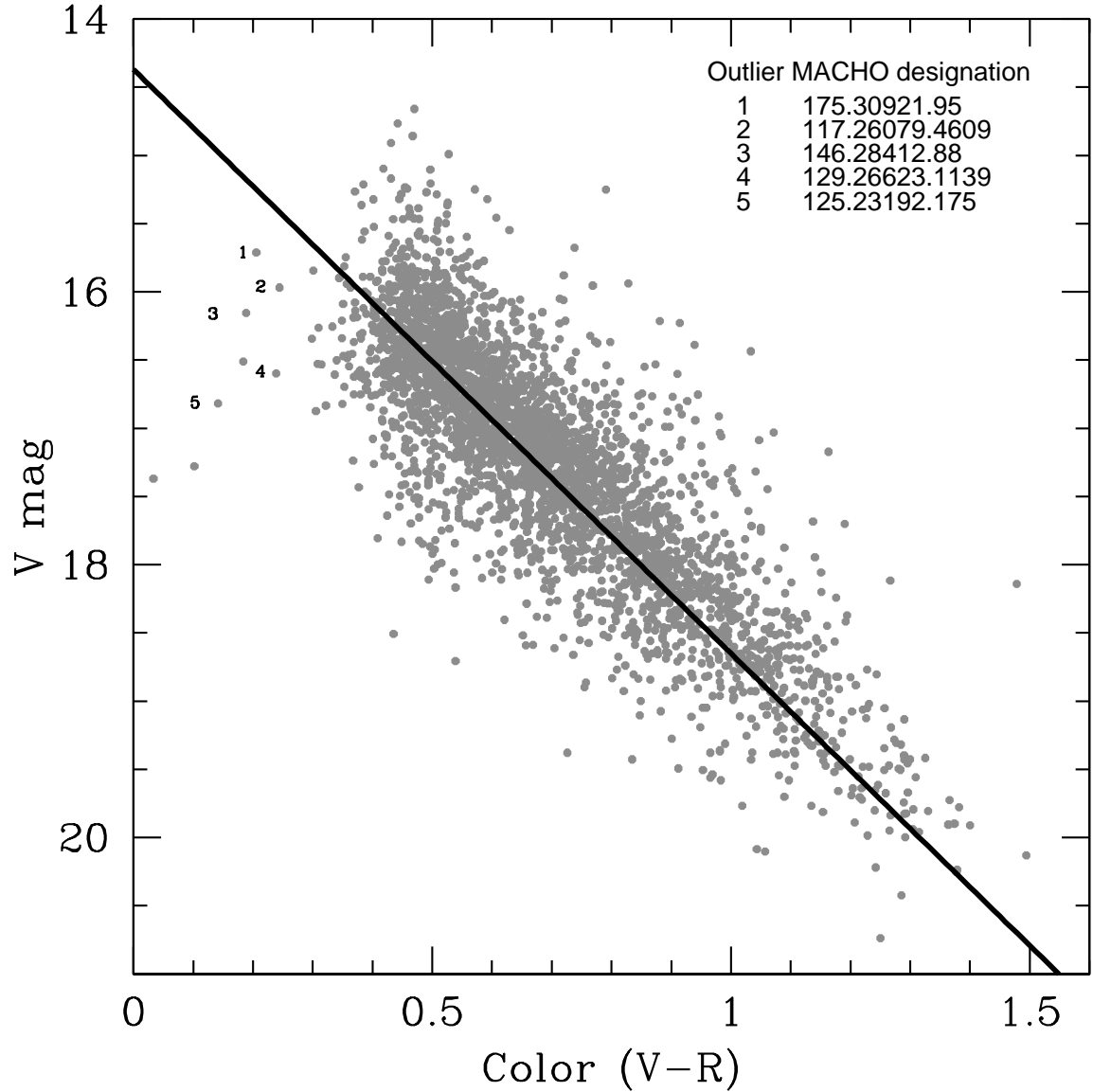}
\caption{Color-magnitude diagram for the 3256 
MACHO Galactic Bulge RR0 Lyrae stars.
The best fit mean $V$ versus mean $(V-R)$ 
relation is shown by the solid line:
    $V_{RR} = (4.28 \pm 0.04)(V-R)_{RR}+ (14.37 \pm 0.03).$
and was obtained using the 
OLS-bisector method \citep{feigel},.
\label{plotone}}
\end{figure}

Figure~\ref{plotone} is a plot of mean $V_{RR}$, versus mean $(V-R)_{RR}$.  
Artificial star tests from \citet{alc99} indicate that the 
limiting magnitude of the photometry 
in the $V$ and $R$ band is about 21.5, which is 
0.5 magnitudes fainter than the faintest RR Lyrae star in this MACHO Bulge 
sample.  The Monte Carlo analysis discussed later in this section
indicates the limiting magnitude has little effect on this paper's
analysis.  

The sources contributing to the intrinsic dispersions of the average $V$ 
and $R$ apparent magnitudes are now
discussed and quantified following the approach from \citet{clem03} .  

1.  photometric errors.  Since many MACHO fields partly overlap on the sky, 
some stars may be counted in the database twice. Using 184 double-represented
stars, the internal photometric uncertainty in $(V-R)$, $V$, and $R$ 
was found by comparing colors and magnitudes evaluated in different fields.
For the MACHO data toward the Galactic Bulge,
the individual error in $V_{RR}$ is $\pm$ 0.15 mag and the individual
error in $(V-R)_{RR}$ is 0.04 (Cook et al.\ 2008, in preparation).
\footnote{For more 
information on the calibration of the MACHO Photometry
RR Lyrae stars Database, see \citet{alc99}, especially pgs. 1547 and 1551.}
As the error in $V_{RR}$ and $(V-R)_{RR}$ are correlated, 
upon multiplying the error in $(V-R)_{RR}$, (0.04 mags), 
by the slope of the fit, (4.28), this value is added to the error
in $V_{RR}$.  The photometric error in the points in Figure~\ref{plotone}
is thus 0.32 mags.

2.  absolute magnitude of RR Lyrae stars. 
It is generally assumed that the absolute magnitude of an RR Lyrae star has 
a linear dependence on $[Fe/H]$, and that $\Delta M_V(RR)$ / $\Delta[Fe/H] 
=$ 0.2 mag $\rm dex^{-1}$.  The dispersion around the mean value of 
$\rm [Fe/H]$ = -1.00 dex found in Galactic Bulge RR Lyrae stars is
$\sigma_V=0.29$ mag. (Kunder et al.\ 2007, in preparation).  This corresponds
to an absolute magnitude dispersion of about 0.06 mag.

The level of evolution off the zero-age horizontal branch (ZAHB) 
can also affect at RR Lyrae star's absolute magnitude.  
From the vertical height of the HB of a number of globular clusters of 
different metallicities, the dispersion around the average RR Lyrae 
luminosity due to the evolution off the ZAHB of each individual RR Lyrae 
is estimated  as $\sim$ 0.08 mag  \citep{clem03, sandage90} .

The scatter in RR Lyrae absolute magnitude due to evolutionary phase, 
helium content, and alpha-element abundance [$\alpha$/Fe], are less
significant and are not accounted for.  

3.  bulge distance spread.

Using the standard Hernquist model density profile with a scale length,
characteristic density, and characteristic velocity of the bulge
from \citet{widrow05} Table~2, the size of the bulge is $\sim$ 1 kpc.
As the galactic latitude and longitude of the MACHO bulge fields ranges 
from -1.5$^\circ$ to - 10$^\circ$ and 0$^\circ$ to 10$^\circ$, respectively,
a correction between the line of sight of the RR Lyrae stars in the bulge 
fields and the plane of the galaxy is negligible.  A distance spread of 1 kpc
corresponds to a $\Delta V$ of $\sim$ 0.29 mags.   

Adding in quadrature all dispersion contributions, the scatter in $V$ for
a fixed $(V-R)$ is 0.44 mags, with the major source of this scatter resulting 
from the extent of the bulge along the line of sight.

Fitting a straight line to the data must be done with caution, since
errors are present in both independent and dependent variables and are
furthermore correlated with each other.  Ordinary least squares (OLS),
which assumes no error in the independent variable, can therefore
not be used here.  Instead a straight line fit to the data is obtained 
using an OLS bisector fit to the individual data points.

The OLS-bisector method has been used when both variables 
are subject to measurement 
error and it is not clear which variable should be treated as the independent 
and which as the dependent.  It has been shown to outperform other
approaches in such cases \citep{babu, isobe90}.  Using the SLOPES program 
\citep{feigel}, the data are fit using the OLS-bisector method.  
The slope found is $4.28 \pm 0.04$ and is shown as a solid line in 
Figure~\ref{plotone}.  Upon restricting the fit to 
those stars with  $(V-R) > 0.34$, a slope of $4.34 \pm 0.04$ is obtained.  
A further cut to remove the very red stars with $(V-R)> 1.1$, 
yields a slope of $4.49 \pm 0.05$.  These fit values are given in
Table~\ref{tab:RvTab}; the $R_{V,VR}$
chosen as the optimal regression is $R_{V,VR} =$ 4.28, as it accounts 
for all data values.  Table~\ref{tab:RvTab} shows that the slope is quite 
sensitive to just a few points, although these formal errors for the 
OLS-bisector method are quite small.  The error in $R_{V,VR}$ of 
$\pm 0.2$ is chosen to encompass the range of slopes.

The OLS-bisector fit to our data is further investigated with a simple
Monte Carlo analysis.  To duplicate the observed RR0 Lyrae data in this
paper, 3500 $E(V-R)$ values between 0 and 1.5 are randomly selected.  
$E(V-R)$ is converted to Av using a selective extinction 
coefficient, $R_{V,VR}$, which we vary.  
Assuming an RR Lyrae absolute magnitude 
of $M_V$ = 0.6 and a distance to the Bulge of 8 kpc, an observed 
$V$ magnitude is found.  As the observed
$V$ magnitudes in the MACHO data have an error of $\sim$ 0.5 mags, 
Gaussian random numbers are generated with a dispersion of 0.5 mags, and 
added to the simulated $V$ magnitudes.  Using $(V-R) = E(V-R) + 0.28$,
and assuming the uncertainty in the measured $(V-R)$ colors is 0.05 magnitudes,
the simulated V magnitudes can be plotted as a function of simulated (V-R) 
magnitudes.  We make sure this simulated plot looks similar to 
Figure~\ref{plotone}.

Next a cut is made to remove all stars which have V and
R magnitudes less than 21.5.  The OLS-bisector method is then used to derive
the slope of the fit before and after the magnitude cut.  We find there is no 
significant change in the value of the slope after the cut.  However, the
slope found using the OLS-bisector method is not always the value of the
slope that was input.  This has implications on the derived $R_{V,VR}$ 
value.  Apparently the errors in the photometry in this paper are sufficient 
to throw off the OLS-bisector fit for a given value.

\begin{figure}
\epsscale{1.0}
\includegraphics[width=16cm]{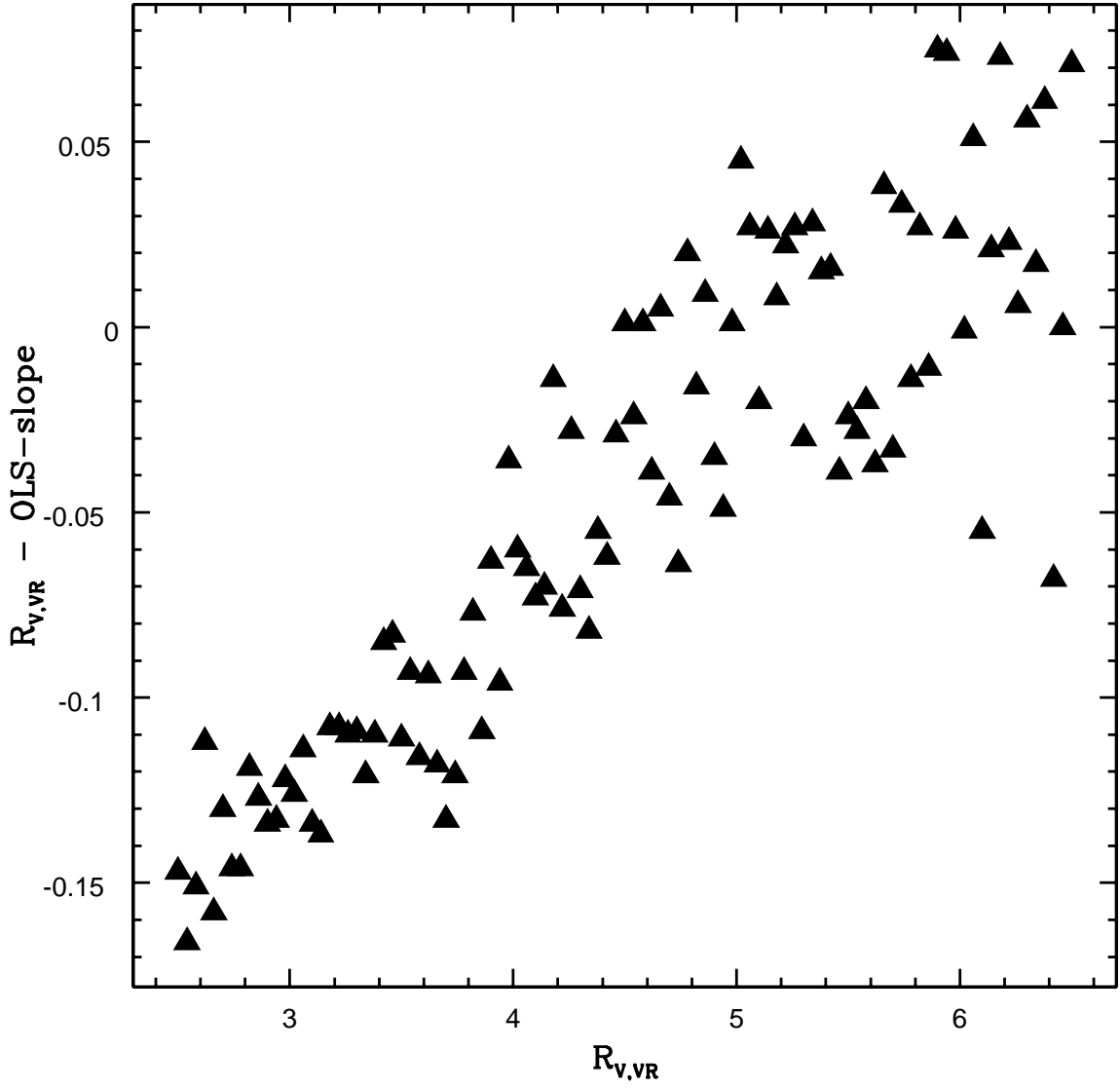}
\caption{
The difference in the true $R_{V,VR}$ value and the OLS-bisector slope, 
$R_{(V,VR)}$ - OLS-fitted slope, as a function 
of the true $R_{V,VR}$ value.
\label{plottwo}}
\end{figure}

Figure~\ref{plottwo} shows the difference in the true $R_{V,VR}$ value
and the OLS-bisector slope,  $R_{(V,VR)}$-OLS slope, as a function 
of the true $R_{V,VR}$ value.  The scatter in this plot is mainly
due to the errors in the $V$-magnitudes and $(V-R)$ colors.  A slope with 
a value of $\sim$ 4.6 exhibits the least 
amount of deviation between the true and calculated $R_{V,VR}$ value.  
The $R_{V,VR}$ value found in this paper is close to 4.6, so the slope
found by the OLS-bisector method should not be significantly biased.  Even
values well above and below 4.6 do not differ from the true $R_{V,VR}$
value by more than 0.2 mags.  Other simulations
with somewhat different input parameters (e.g., uncertainty in 
photometry) are performed and results similar, but not identical to 
Figure~\ref{plottwo}, are found.  Thus, we can not simply correct the slope
that is obtained from the real data using these simulations.  Thus,
an uncertainty of 0.2 in the $R_{V,VR}$ is adopted.

\subsection{Comparisons and Consistency}
It is an interesting question how this $R_{V,VR}$ value from the MACHO
RR Lyrae stars compares with other reddening relations.  Reddening
relations determined from broad band photometric data depend on the 
intrinsic spectral energy distributions of the objects used to
determine them as well as the particular filters. The MACHO $b_M$ and $r_M$
filters are considerably different from Johnson $V$ and Kron-Cousins $R$.

To compare our results to the standard extinction
curve, the \citet{fitz99} extinction curve with a 
constant total-to-selective extinction parameter, $R_{V,BV} = 3.1$, 
is used with the PHOENIX model synthetic spectra \citep{hauschildt99}.
The Johnson BVR transmission curves from \citet{bessell90}
yield $R_{V,VR} = 4.88$ for $T_{eff}=6000$ K, 
$R_{V,VR} = 4.92$ for $T_{eff}=7000$ K and 
$R_{V,VR} = 5.07$ for $T_{eff}=30000$ K.  An effective temperature of 
6000 K - 7000 K is typical of RR Lyrae stars, and an effective temperature of
30000 K is typical of early type stars.  
The synthetic spectra adopted is one with $\rm [Fe/H] = -1.0$ and 
$\rm log g = 2.5$,
but varying $\rm [Fe/H]$ and surface gravity does not significantly
affect the $R_{V,VR}$ values.  

The MACHO $b_M$ and $r_M$ filters are non-standard, but were transformed into
the Johnson $V$ and Kron-Cousins $R$ as described in \citet{alc99}.  
We simulated this procedure by using the MACHO $b_M$ and $r_M$ transmission
curves and our synthetic spectra to obtain MACHO $b_M$ and $r_M$ magnitudes. 
These were then converted using the \citet{alc99} transformation.
It is found that $R_{V,VR} = 4.34$ for $T_{eff}=6000$ K, and 
$R_{V,VR} = 4.26$ for $T_{eff}=7000$ K.  Both these values
are in agreement with the value obtained in this analysis. In comparison, 
a $R_{V,BV}$ = 2.5 would yield $R_{V,VR} = 3.77$.

Reddening laws are normally derived from differential measurement of 
reddened and unreddened luminous early-type stars.  But such stars are 
absent in the Galactic Bulge.  This analysis uses Population II type stars 
and indicates that the reddening law in the bulge is on average similar
to the standard solar neighborhood value of $R_{V,BV} = 3.1$.
Changing  $R_{V,BV}$ to e.g. a non-standard value, does significantly 
affect the measured $R_{V,VR}$.  Quantitatively, it is found that the 
change in $R_{V,VR}$ is 68\% of the change in $R_{V,BV}$.

Using Galactic bulge Red Clump Giants from the OGLE dataset, 
\citet{udalski03} present substantial evidence that
the ratio of total to selective absorption, $R_{V,VI}$, is much
smaller toward the Galactic Bulge than the value corresponding
to the standard extinction curve and that $R_{V,BV}$ varies considerably
along different lines of sight.  Their value of $R_{V,VI} \sim 2.1$,
as opposed to the standard value of about 2.5, corresponds to
an $R_{V,BV}$ of about 2.6.  
\citet{stan96} constructed extinction and reddening maps for Baade's 
window from color-magnitude diagrams obtained by OGLE which showed that 
extinction in Baade's window is quite irregular, varying between 1.3 and 
2.8 mag in $A_{V,BV}$, with an estimated error of $\sim$ 0.1 mag. 
\citet{ruffle04} use 70 Galactic planetary nebulae observed using 
narrow-band filters to find observed $<R_{V,BV}>$ = 2.0 toward the bulge.  

The $R_{V,VR}$ value determined from the RR Lyrae stars above, 
indicates that the standard Galactic reddening law can in general
be adopted in studies of objects toward the bulge.  We
caution that the value of $R_{V,BV}$ can vary in certain
directions but that on average, the standard reddening law can be used.

\section{Reddening}
\subsection{Determination of intrinsic $(V-R)_0$ colors}
\citet{oldGuy} and \citet{blanc92} argue that apparent $(B-V)$ colors of RR 
Lyrae stars at phases close to the minimum light can be utilized to
measure the amount of interstellar reddening along the line of sight to the 
star since the intrinsic $(B-V)_0$ colors seem constant in the phase
interval 0.5 -- 0.8.  \citet{mateo} suggest
that $(V-I)$ colors might also be used, perhaps with a smaller or 
negligible metallicity correction.  We investigate this further 
for $(V-R)$ colors below.

The colors at minimum light, $(V-R)_0$, can be computed via two different 
methods.  In the first method, the Fourier fits to both $V$ and $R$-band data
are used to average the colors, in magnitude units, for phases
between 0.5 and 0.8.  However, we find that when $(V-R)_0$ is 
determined using this method, it correlated with 
amplitude.  In the second method, the $(V-R)$ color is found at minimum 
$V$-band light. 
\citet{shashi04} used this technique and found that 
intrinsic color at minimum $V$-band light is nearly independent of period
and $V$-band variation amplitude.  Since this method
yields $(V-R)_0$ values that are very weakly correlated with amplitude, 
this is the method employed here. 
Table~\ref{tab:chabTab} presents our determination of
unreddened $(V-R)_0$ colors for eleven well observed field RR0 variables 
(the data in Table~\ref{tab:chabTab} come from 
Liu \& Janes 1989; Cacciari et al.\ 1987; and Layden 2005, private 
communication).  
The color excess, $E(V-R)$, is computed by multiplying the published $E(B-V)$
values by the ratio of $R_{V,BV}/R_{V,VR}=3.315/5.16$, where the values 
$R_{V,BV} = 3.315$ and $R_{V,VR}=5.16$ are taken from
\citet{schlegel98}.\footnote{We explicitly assume here that for this sample
of RR Lyrae stars, extinction is on average well represented by the
standard $R_{V,BV}$ value.}
A plot of $E(V-R)$ vs $(V-R)$ has a linear character.  This is
quite remarkable, considering some of these stars exhibit 
the Blazhko effect (e.g., AW And), some are affected by both the Blazhko 
effect and a "phase-lag" problem (e.g., SS For), and some
have evidence of shock waves in their atmospheres (e.g., RV Phe, V440 Sgr)
as discussed by \citet{cacciari89}, \citet{cacciari289}.
It has been shown that in many RR Lyrae stars the 
Blazhko effect does not affect colors at minimum light, making this
method robust.  We conclude that the mean $(V-R)_0$ color of RR0 Lyrae stars at
minimum $V$-band light is $0.28 \pm 0.02$.  
Table~\ref{tab:chabTab} suggests that the intrinsic
color at minimum light is independent or at least very insensitive
to metallicity.

\subsection{Reddening toward the bulge RR Lyrae stars}
The {\em observed} and {\em intrinsic} colors at minimum $V$-band
are put forth to determine the color excess of the bulge RR0 Lyrae stars: 
\begin{equation}
E(V-R) = (V-R) - (V-R)_0.
\end{equation}

The Fourier fits performed to find $(V-R)$ color at minimum $V$-band light
are sensitive to the quality of a lightcurve. 
To insure high $E(V-R)$ accuracy without sacrificing the sample size,
the Fourier fits in both passbands for all stars are individually 
examined.  Only those with fits that approximated the RR Lyrae
light curve nicely are taken (i.e., if the Fourier method mimicked features
in the RR Lyrae lightcurve well, such as the dip and sharp rise of the 
curve at minimum light).
This criterion results in 3525 RR0 Lyrae stars with well determined
reddening values. Example entries of this reddening catalog are presented
in Table 3; the complete listing is available in the electronic version.

\begin{figure}[htb]
\includegraphics[width=16cm]{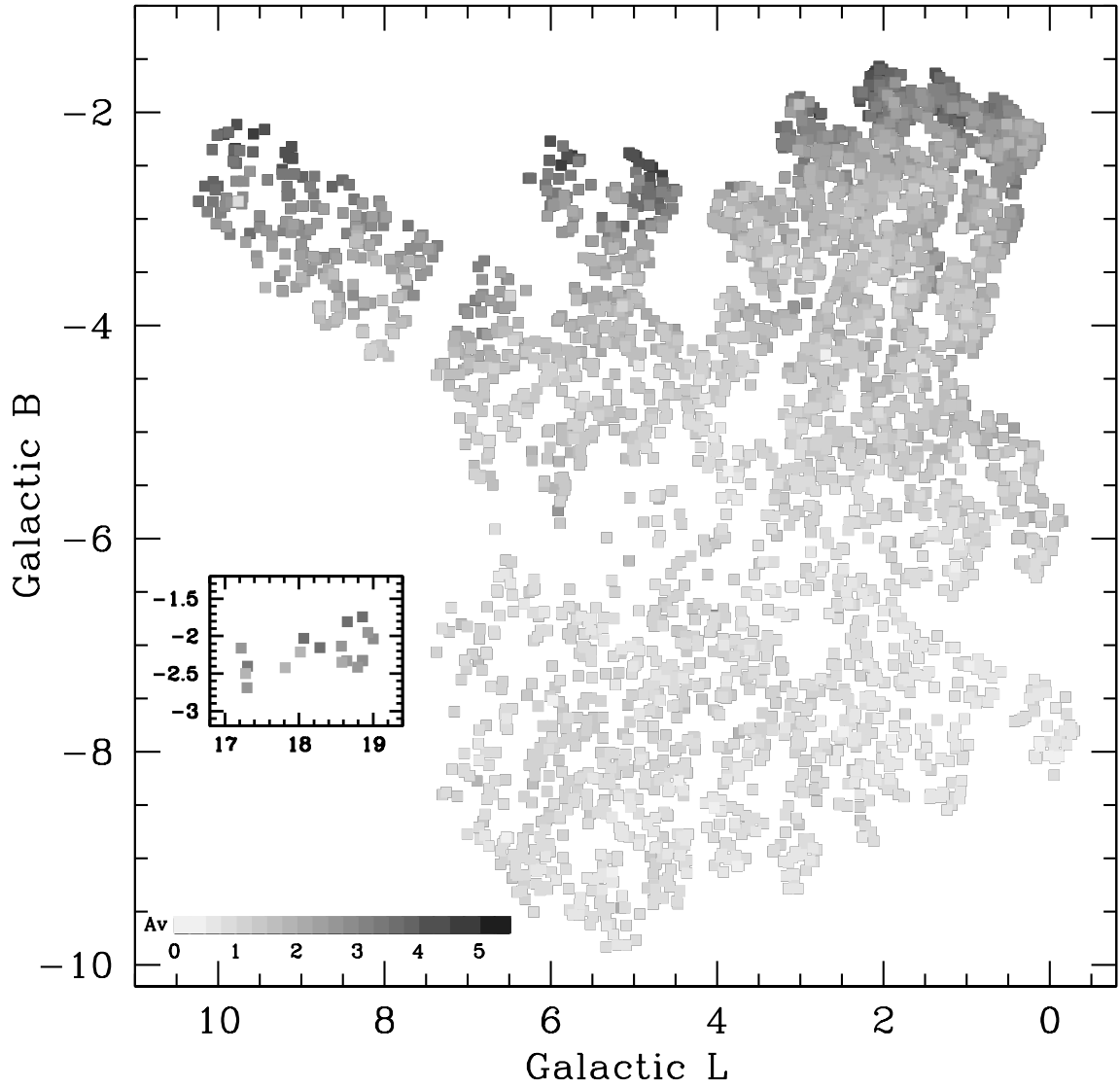}
\caption{Reddening map of $A_V$ values of the central Galactic region.  
The scale is given at the bottom.  Insert presents three disk fields separated
from the others by 10 $\!\!^\circ$.  
\label{plotthree}}
\end{figure}

Multiplying the 3525 $E(V-R)_{RR}$ values by the selective extinction 
coefficient, $R_{V,VR}$, allows the determination of the visual extinction.
A map of the visual extinction, $A_V$, is shown in Figure~\ref{plotthree}.
As \citet{pop03} also remark, it is immediately obvious that on large 
scales, extinction is regularly stratified parallel to the Galactic plane.

To determine the distance modulus to the Galactic Bulge from the RR0 Lyrae
stars, an RR Lyrae absolute magnitude of $\rm M_V=0.59\pm 0.03$ at 
$\rm [Fe/H]=-1.5$ as compiled by \citet{cacc03} is adopted.
If a metallicity dependence of $\rm M_V=(0.25\pm 0.05)[Fe/H] + constant$ is
adopted (consistent with the range of slopes found in the literature), 
then using the average metallicity of $\rm [Fe/H]=-1$ from above,
an $\rm M_V = 0.72\pm 0.04$ for RR Lyrae stars in the Bulge is obtained. 
This predicts $\rm (m-M)_0 = 14.7\pm 0.1$ for Bulge RR Lyrae.
Adopting the RR Lyrae statistical parallax calibration 
of $\rm M_V=0.77\pm 0.13$ at $\rm [Fe/H]=-1.6$ from \citet{goul98}, then using 
the same metallicity dependence, an $\rm (m-M)_0 = 14.5\pm 0.16$ for RR Lyrae 
stars in the Galactic Bulge is predicted.  This is identical to the
Galactocentric distance modulus measured by \citet{eisen03} from
the orbit of a star around the central black hole.

\subsection{Comparison with complementary MACHO determination}
\begin{figure}[htb]
\includegraphics[width=16cm]{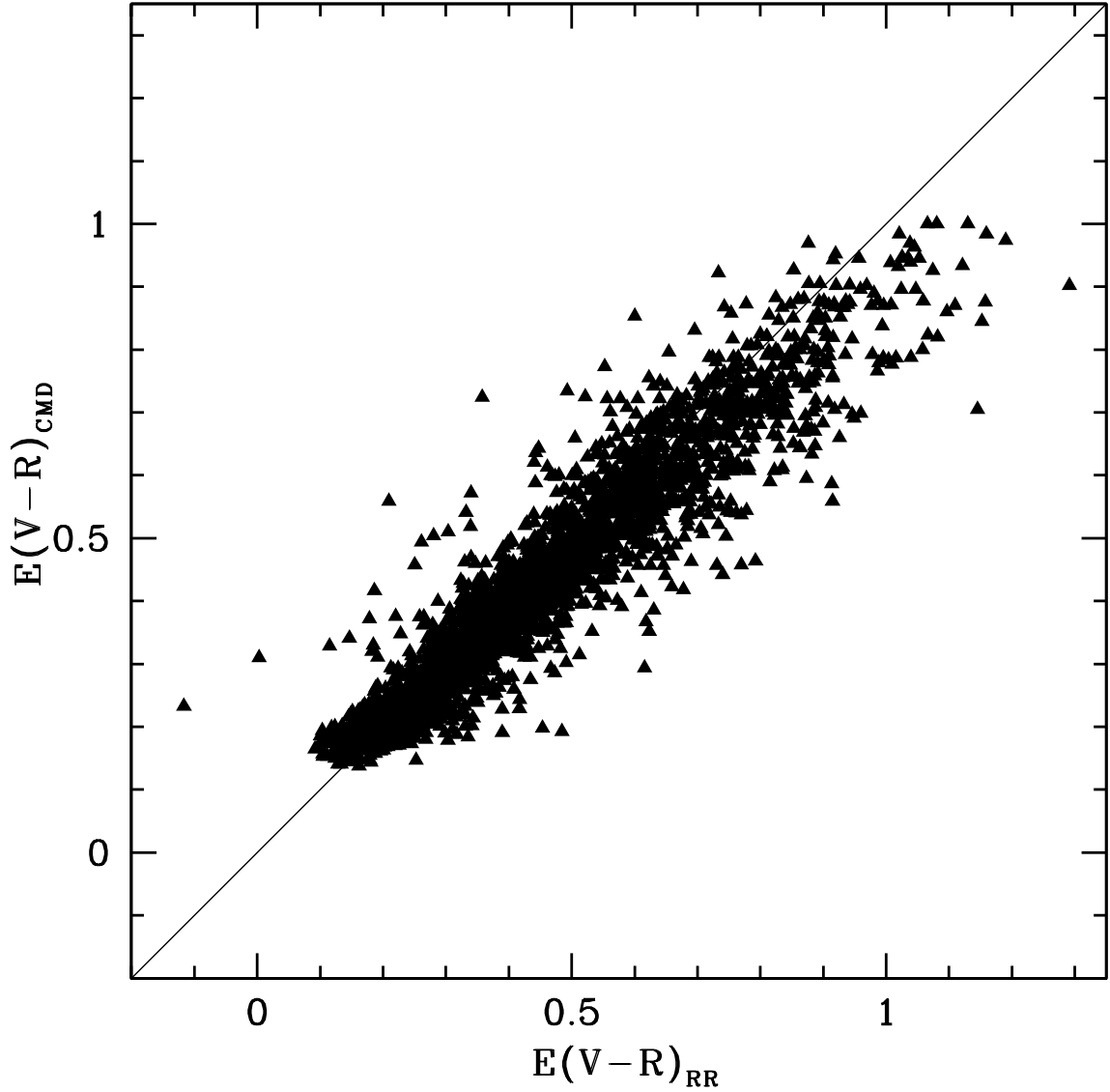}
\caption{A comparison of $E(V-R)$ derived using $(V-R)$ at minimum
light, $E(V-R)_{\rm RR}$, for 2605 RR Lyrae stars with color-magnitude
based  $E(V-R)_{\rm CMD}$ from \citet{pop03}.
The relation is linear, with $E(V-R)_{\rm RR}$ values
slightly larger than $E(V-R)_{\rm CMD}$ at high extinction.
\label{plotfour}}
\end{figure}

In this subsection the reddenings obtained here are tested 
against \citet{pop03} values,
which are based on photometry from the MACHO bulge fields.
The MACHO data of the Galactic Bulge are based on observations of 94
bulge fields, each field covering an area of 43' by 43'. \citet{pop03}
use 4' x 4' sky regions (tiles) as
resolution elements.  They show that mean colors for these tiles can
be converted to extinction, and can thus be used to derive a visual 
extinction map.  Each RR0 Lyrae star is matched with the corresponding
tile, and the color excess, $E(V-R)_{\rm RR}$, of the line of sight 
to the RR0 Lyrae star is compared to the \citet{pop03}, $E(V-R)_{\rm CMD}$,  
as calibrated to \citet{stan96} extinction\footnote{The \citet{pop03}, 
$E(V-R)_{\rm CMD}$, as calibrated to \citet{dutra} extinction
is about $0.07$ redder than $E(V-R)_{\rm CMD}$ calibrated to \citet{stan96}.}.
Figure~\ref{plotfour} shows this linear relationship, 
with the line of slope unity over-plotted.  The $E(V-R)_{\rm RR}$ values
become larger than $E(V-R)_{\rm CMD}$ when reddening increases, which 
could mean the \citet{pop03} extinction map 
underestimates the extinction in areas with high reddening.
Performing the \citet{bessell99} calibration to the MACHO RR Lyrae
lightcurve data does not alter this result.
An increase in scatter is expected in stars with high 
reddenings, since 
generally a redder RR0 Lyrae star is fainter and is more likely to have noisy 
photometry and lightcurves.  
Especially near the Galactic center, differential reddening due to nonuniform 
distribution of the intervening dust on small spatial scales makes the 
study of the stellar population difficult even in the infrared 
\citep[see, e.g.,][]{nara96,davidge98}. 
\citet{frogel99} show in their Figure~5, that the amount of differential 
reddening in a field is directly proportional to the average reddening:
the more reddening present in a field, the more patchy that reddening is. 
As the \citet{pop03} extinction map is based on the average color of 
4' x 4' sky regions, their average colors and hence reddenings would
be susceptible to differential reddening, explaining the slight discrepancy
between their reddenings and ours at higher reddenings.

What could cause the systematic bias in the $E(V-R)_{\rm RR}$ values 
to become larger than the 
$E(V-R)_{\rm CMD}$ as the reddening increases?  
\citet{pop03} perform a calibration to \citet{stan96} extinction and this 
calibration assumes a uniform $R_{V,BV}$ value.  This assumption
does not hold, however, since the extinction toward the Bulge
has nonstandard properties (Popowski 2000).  
\citet{udalski03} find that the difference between the 
measured line of sight $R_{V,VI}$, and the standard value of $R_{V,VI}$
depends on color.  Furthermore, they find that this difference
is larger for redder $V-I$ colors and practically negligible in the range
where the OGLE data were calibrated by standards $(V-I<2)$.  
Similarly, the difference in the coefficient of selective 
extinction $R_{V,VR}$ and the standard $R_{V,VR}$ would also increase 
with increasing $V-R$ and hence with increasing $E(V-R)_{CMD}$ (since 
\citet{pop03} uses the average color to determine $E(V-R)_{CMD}$).  
As the \citet{pop03} $E(V-R)_{CMD}$ calibration neglects a change in 
$R_{V,VR}$ with increasing color, 
at higher $E(V-R)_{CMD}$ the values will be underestimated.

\subsubsection{Outliers}
The most obvious outlier in Figure~\ref{plotfour}
is the RR0 Lyrae star that is designated in the
MACHO database as 175.30921.95. It has a negative $E(V-R)_{RR}$, 
which is unphysical. It has a period, 
amplitude and light curve
shape that confirm it is a fundamental mode pulsator.  However, as is clear 
from its position on a color-magnitude diagram, with $(V-R) = 0.21$
it is one of the bluest stars in the dataset (see Figure~\ref{plotone}).  
\citet{oldGuy} suggests that as many as 3\% of the RR0 Lyrae stars may have 
an unusual blueness associated with them.  The only stars with a color 
bluer than this star and with an R-band lightcurve suitable for reddening
determination, are stars 125.23192.175 and 117.26079.4609, with
$(V-R) = 0.14$ and $(V-R) = 0.19$, respectively.  The bluer of these
two stars' $E(V-R)_{RR}$ deviates by 0.3 mag from $E(V-R)_{CMD}$, while the
other one by only 0.06 mag. Other very blue stars include 
129.26623.1139 with $(V-R) = 0.24$ and 146.28412.88 with $(V-R) = 0.25$, 
but their $E(V-R)_{RR}$ is in agreement with
$E(V-R)_{CMD}$.  One more characteristic that distinguishes the blue
outlying stars 175.30921.95 and 125.23192.175 from their blue
non-outlying counterparts, is the fact that
these stars have periods that lie in the range 0.469-0.472 days, 
a range in which unstable light curves are common \citep{kinman92}. 
Spectra of unusually blue stars in the period interval of 0.469-0.472 days,
AR Her and BB Vir, were taken by \citet{oldGuy} and \citet{fernley96}.  
It was found that these stars have systematically 
stronger hydrogen lines at minimum light than all the other stars, indicating
a higher temperature.  The cause of the higher temperature, however, 
remains unknown.  \citet{fernley96} suggested
that BB Vir could be a RR0 Lyrae star that for some reason has evolved a long 
way beyond the fundamental blue edge without yet changing mode.  The 
``hysteresis'' effect is well-known in globular clusters and it is possible 
that stars 175.30921.95 and 125.23192.175 are also extreme examples of it.  
Stars at the blue edge of the instability strip should have a larger amplitude 
\citep{sandage81}.
Stars 175.30921.95 and 125.23192.175, the blue stars with deviating 
$E(V-R)$ values, have about twice as large $V$ amplitudes
as the other two blue stars, 129.26623.1139 and 146.28412.88.  It is also 
possible that the MACHO photometry of the star 
175.30921.95 and 125.23192.175 is incorrect. Follow-up
observations would help clarify this point.  In conclusion,
although rare, reddening values for RR0 Lyrae stars
determined by finding the $(V-R)$ color at 
minimum $V$-band light can be in error if the star is unusually blue.  This can
be corrected if the temperature of the RR0 Lyrae star is known. 

\section{Conclusions}

A quantitative analysis of reddening toward the Galactic Bulge is performed.  
Assuming that the extinction affects the $V$ magnitudes more 
significantly than 
the distance spread and the intrinsic scatter in absolute magnitudes,
we interpret the slope of the stellar locus on the $V$ versus
$(V-R)$ diagram as the selective extinction coefficient, 
$R_{V,VR}=A_V/E(V-R)$.
From the MACHO RR0 Lyrae sample toward
the Galactic Bulge, we derive $R_{V,VR}=4.3 \pm 0.2$.
This corresponds to the average
value observed in the solar neighborhood $R_{V,BV} = 3.1 \pm 0.3$, and
indicates that the optical reddening law toward the bulge 
is consistent with the standard extinction law.  We show that
the mean minimum-light $(V-R)_0$ color of an RR0 Lyrae star
is nearly constant at minimum 
$V$-band light and equals $0.28 \pm 0.02$, regardless of metallicity.  
Using this property, reddening values to 3525 
RR0 Lyrae stars are found.  Parameters derived for individual stars are 
presented in the electronic version.

\acknowledgments

We would like to thank the referee, Derck Massa, whose thorough report 
has led to substantial improvements to this paper.
Andrea Kunder thanks the Max-Planck-Institut f\"{u}r Astrophysik and Lawrence
Livermore National Laboratory, where
part of this work was completed, for their wonderful people and hospitality,
especially Martin Jubelgas, David Syphers, and Andreas Hamm.
KHC's and part of AMK's work was performed under the auspices of the U.S.
Department of Energy by Lawrence Livermore National Laboratory in part
under Contract W-7405-Eng-48 and in part under Contract DE-AC52-07NA27344.

\clearpage

\begin{table}
\centering
\caption{ Slope ($R_{V,VR}$) of 3256 Bulge RR0 stars plotted on a $V$
  versus $(V-R)$ CMD.}
\label{tab:RvTab}
\begin{tabular}{lcccccc} \\ \hline
Fit&Slope&Error&zero point&Error\\ \hline
OLS bisector, all stars & 4.28 & 0.04 & 14.37 & 0.03\\ 
OLS bisector,  $(V-R)  >$ 0.34 & 4.34 & 0.05 & 14.32 & 0.03\\
OLS bisector, $(V-R) > $0.34, $(V-R)<$1.11 & 4.48 & 0.05 & 14.24 & 0.03\\
\hline
\end{tabular}
\end{table}

\begin{table}
\centering
\caption{Derivation of minimum-light $(V-R)_0$ colors of field RR0 Lyrae 
stars}
\label{tab:chabTab}
\begin{tabular}{lcccccc} \\ \hline
Star&$(V-R)$&$E(V-R)$&$(V-R)_0$&[Fe/H]&$V$ Amp\\ \hline
SWAnd & 0.34 & 0.035 &  0.295 & $-0.24$ & 0.89\\
RRCet & 0.30 & 0.019 &  0.301 & $-1.45$ & 0.78\\ 
RXEri & 0.34 & 0.053 &  0.277 & $-1.33$ & 0.87\\
RRGem & 0.35 & 0.054 &  0.276 & $-0.29$ & 1.15\\
RRLeo & 0.31 & 0.034 &  0.266 & $-1.60$ & 1.23\\
TTLyn & 0.30 & 0.014 &  0.286 & $-1.56$ & 0.70\\
AVPeg & 0.34 & 0.056 &  0.274 & $-0.08$ & 1.08\\
TUUMa & 0.30 & 0.018 &  0.272 & $-1.51$ & 1.02\\
SSFor & 0.27 & 0.012 &  0.248 & $-0.94$ & 1.23\\
RVPhe & 0.30 & 0.006 &  0.284 & $-1.69$ & 0.71\\
V440Sgr & 0.33 & 0.076 & 0.254 & $-1.40$ & 1.16\\
\hline
\end{tabular}
\end{table}

\begin{table}
\begin{scriptsize}
\centering
\caption{Fragment of the catalog of reddenings for 3525 RR0 Lyrae stars
in the MACHO Galactic bulge fields}
\label{tab:header}
\begin{tabular}{lcccccccccc} \\ \hline
Star ID & Ra & Dec & Gal $l$ & Gal $b$ & Period & $V$ mag & $R$ mag & $V$ amp & $(V-R)_{min}$ & $E(V-R)$\\ \hline
101.20648.645 & 18:04:31.80 & -27:33:55.8 & 3.286 & -2.904 & 0.590062 & 17.58 & 16.77 & 0.57 & 0.808 & 0.53\\
101.20649.500 & 18:04:21.36 & -27:27:44.6 & 3.357 & -2.820 & 0.570140 & 17.11 & 16.46 & 0.70 & 0.729 & 0.45 \\
101.20650.531 & 18:04:25.32 & -27:26:16.1 & 3.385 & -2.821 & 0.557547 & 17.29 & 16.44 & 0.81 & 0.723 & 0.44 \\
101.20650.965 & 18:04:24.24 & -27:27:40.7 & 3.363 & -2.829 & 0.505202 & 18.38 & 17.57 & 0.94 & 0.829 & 0.55 \\
101.20652.684 & 18:04:23.16 & -27:17:18.2 & 3.512 & -2.740 & 0.570760 & 17.07 & 16.36 & 0.25 & 0.720 & 0.44 \\
101.20653.526 & 18:04:25.32 & -27:11:42.4 & 3.597 & -2.702 & 0.469922 & 17.26 & 16.53 & 0.76 & 0.688 & 0.41 \\
101.20654.508 & 18:04:30.72 & -27:11:14.6 & 3.614 & -2.716 & 0.552941 & 17.13 & 16.33 & 0.88 & 0.824 & 0.54 \\
101.20654.938 & 18:04:32.88 & -27:08:09.6 & 3.663 & -2.697 & 0.457945 & 17.51 & 16.78 & 0.70 & 0.778 & 0.50 \\
101.20778.1172 & 18:04:36.84 & -27:35:11.4 & 3.276 & -2.930 & 0.514020 & 17.76 & 16.95 & 0.99 & 0.817 & 0.54 \\
101.20779.378 & 18:04:41.16 & -27:28:17.4 & 3.385 & -2.888 & 0.633424 & 17.06 & 16.26 & 0.53 & 0.835 & 0.55 \\
\hline
\end{tabular}
\end{scriptsize}
\end{table}

\clearpage

\clearpage

\clearpage
\end{document}